\documentclass[prd,preprint,nofootinbib]{revtex4}
\usepackage[dvipdfmx]{graphicx}
\usepackage{amsfonts}
\usepackage{latexsym}
\usepackage{amsmath}
\usepackage{amssymb}
\usepackage{epsfig}
\begin{document}

\title{Smooth hybrid inflation in light of ACT DR6 data }

\author{Nobuchika Okada}
 \email{okadan@ua.edu}
 \affiliation{Department of Physics and Astronomy, University of Alabama, Tuscaloosa, Alabama 35487, USA}

\author{Osamu Seto}
 \email{seto@particle.sci.hokudai.ac.jp}
 \affiliation{Department of Physics, Hokkaido University, Sapporo 060-0810, Japan}

%

\begin{abstract}
Smooth hybrid inflation is a hybrid inflation model that is free from topological defects
 and predicts the density perturbation with the spectral index of about $0.97$.
We show that the prediction on the spectral index is robust
 regardless the power of nonrenormalizable terms and 
 meets with the latest results reported by Atacama Cosmology Telescope.
Viable scenarios for baryogenesis and dark matter are also discussed.
\end{abstract}

\preprint{EPHOU-25-011} 

\vspace*{3cm}
\maketitle


\section{Introduction}

The inflationary cosmology has been the standard paradigm of modern cosmology.
It solves various problems in the standard big bang
 cosmology~\cite{Starobinsky:1979ty,Sato:1981qmu,Guth:1980zm,Linde:1981mu,Albrecht:1982wi}
 and explains the origin of the density perturbation in our Universe from
 the quantum nature of an inflaton field~\cite{Hawking:1982cz,Starobinsky:1982ee,Guth:1982ec}.
The scalar perturbation is generated from the inflaton's quantum fluctuation
 during quasi-de Sitter inflationary expansion.
As a result, the spectrum is almost scale invariant, in fact its scale dependence,
 parametrized by the spectral index, reflects the dynamics of inflation. 
The scalar spectral index has been reported as $n_s= 0.965 \pm 0.004$ from the measurement
 by Planck~\cite{Planck:2018vyg}.
Thus, it has been regarded that so called Starobinsky inflation~\cite{Starobinsky:1980te},
 Higgs inflation~\cite{Bezrukov:2007ep} and $\alpha$-attractor
 inflation~\cite{Kallosh:2013hoa,Ferrara:2013rsa,Kallosh:2013yoa}
 in the class of pole inflation~\cite{Galante:2014ifa,Broy:2015qna}
 are promising models of inflation, because those models predict $n_s \simeq 0.96$.

Recently,
 the latest results from Atacama Cosmology Telescope (ACT)~\cite{ACT:2025fju,ACT:2025tim} 
 have been released and the scalar spectral index of $n_s = 0.974\pm 0.003$ has been reported~\cite{ACT:2025fju}
 for after combining data of ACT, Planck, and baryon acoustic oscillation.
This shift of estimated $n_s$ motivates us to reconsider models for inflation~
\cite{Kallosh:2025rni,Aoki:2025wld,Dioguardi:2025vci,Gialamas:2025kef,Salvio:2025izr,Antoniadis:2025pfa,Dioguardi:2025mpp,Rehman:2025fja,Gao:2025onc,He:2025bli,Drees:2025ngb,Yin:2025rrs,McDonald:2025odl,Byrnes:2025kit,Maity:2025czp,Yi:2025dms,Gialamas:2025ofz,Yogesh:2025wak,Haque:2025uga,Pallis:2025nrv,Wolf:2025ecy,McDonald:2025tfp,Choudhury:2025vso}.
Although hybrid inflation models~\cite{Linde:1993cn,Copeland:1994vg} can be constructed
within realistic particle physics models~\cite{Lyth:1998xn}, in particular, in the supersymmetric model~\cite{Dvali:1994ms}, 
 those have appeared not to be promising, because those generally predict a too large spectral index. 
In addition, hybrid inflation models suffer from the formation of topological defects,
 because the spontaneous symmetry breaking by the waterfall field terminates inflation
 with generating topological defects.
The typical mass per unit length of cosmic strings formed after hybrid inflation
 confronts with the present observational bound on it~\cite{Battye:2010xz,Planck:2013mgr}.

In this article, we revisit smooth hybrid inflation,
 which has been proposed as a hybrid inflation model free from topological defects
 and predicts the spectral index of about $0.97$~\cite{Lazarides:1995vr}.
We show that the prediction on the spectral index is robust
 regardless of the power of nonrenormalizable terms and well meets with the latest ACT DR6 data.

\section{Smooth hybrid inflation}

The superpotential for a smooth hybrid inflation is given by
\begin{equation}
 W = S\left(-\mu^2+\frac{(\bar{\Psi}\Psi)^m}{M^{2m-2}}\right),
\end{equation}
 where this model is based on certain gauge group $G$ under which $S$ is a singlet,
 and $\Psi$ and $\bar{\Psi}$ are a pair of a vectorlike representation. 
 We may consider $U(1)_{B-L}$ as an example. 
It is of the most general form consistent with $R$ symmetry 
under which 
$(S, \Psi, \bar{\Psi}) \rightarrow (e^{i2\alpha} S,e^{i2\alpha}\Psi,e^{-i2\alpha} \bar{\Psi})$,
 $W \rightarrow e^{i2\alpha} W$ while $\bar{\Psi}\Psi$ is invariant
 and posseses a discrete symmetry under which $\bar{\Psi}\Psi$ changes sign or phase.
While the $m=2$ case has been considered and derived by imposing two $Z_4$ symmetries
 in the original paper~\cite{Lazarides:1995vr} and by imposing an extra $Z_2$ symmetry $\Psi \rightarrow -\Psi$ in Ref.~\cite{Jeannerot:2001qu}, we generalize 
 that $m$ is an integer ($\geq 2$) as in Refs.~\cite{Yamaguchi:2004tn,Kawasaki:2006zv}
 for other possible realizations.
With the minimal Kahler potential, the scalar potential in supergravity in the Planck unit is 
\begin{align} 
V =&
 e^{|S|^2+|\Psi|^2+|\bar{\Psi}|^2}\left[
 \left|-\mu^2+\frac{(\bar{\Psi}\Psi)^m}{M^{2m-2}}\right|^2 (1-|S|^2+|S|^4 )
 \right. \nonumber \\
 &
 +|S|^2 \left|m\bar{\Psi}\frac{(\bar{\Psi}\Psi)^{m-1}}{M^{2m-2}} 
 +\Psi^* \left(-\mu^2+\frac{(\bar{\Psi}\Psi)^m}{M^{2m-2}}\right) \right|^2  
 \nonumber \\
 & \left. 
 +|S|^2 \left|m\Psi\frac{(\bar{\Psi}\Psi)^{m-1}}{M^{2m-2}} 
 +\bar{\Psi}^* \left(-\mu^2+\frac{(\bar{\Psi}\Psi)^m}{M^{2m-2}}\right) \right|^2 
 \right] + D {\rm -terms} \nonumber \\
  \simeq& 
 \left(-\mu^2+\frac{\psi^{2m}}{M^{2m-2}}\right)^2 
 \left(1+\frac{|S|^4}{2}+2\psi^2 \right)
 +2|S|^2 \psi^2 \left(m\frac{\psi^{2m-2}}{M^{2m-2}}-\mu^2 \right)^2 .
\label{smooth:potential}
\end{align} 
Here, in the last expression,
 the D-flat condition ($\Psi = \bar{\Psi}^* \equiv \psi$) is imposed, and
 the third terms in the F-term of $\Psi$ and $\bar{\Psi}$ are neglected compared to the first terms,
 and the lowest order supergravity corrections for $S$ and $\psi$ only are kept in the F term of $S$,
 since we are interested in the field region $|S|^2, \psi^2 \ll 1$.
The supersymmetric global minimum is located at 
\begin{equation} 
 \left( S, \psi \right) = \left( 0, \mu^{\frac{1}{m}}M^{\frac{m-1}{m}} \right), 
\label{eq:minimum}
\end{equation}
 where the mass of inflatons is 
\begin{equation} 
 m_{S}^2 = m_{\psi}^2 
 \simeq 2m^2\mu^2\left(\frac{\mu}{M}\right)^{\frac{2(m-1)}{m}}.
\end{equation}

\begin{figure}[htbp]
\centering
\includegraphics[clip,width=11.0cm]{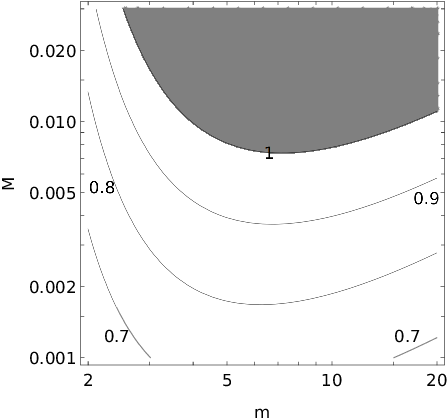}
\caption{
Contours of $\sigma(N)/\sigma_t$ for $N=60$.
The gray shaded region violates the condition (\ref{enq:sigma}). }
\label{Fig:m-M}
\end{figure}

The stationary condition for $\psi$ is
\begin{equation}
V,_{\psi}
 \simeq 4\psi\left(-\mu^2+m\frac{\psi^{2m-2}}{M^{2m-2}}\right)
 \left(-\mu^2+\frac{\psi^{2m}}{M^{2m-2}}
 +m(2m-1)|S|^2\frac{\psi^{2m-2}}{M^{2m-2}}\right) ,
\end{equation}
 and we find a nontrivial local minimum
\begin{equation}
 |S|^2\left(\frac{\psi}{M}\right)^{2m-2}=\frac{\mu^2}{m(2m-1)} ,
\label{S-psi:relation}
\end{equation}
 for a large $|S|$ as
\begin{equation}
|S|^2 > \frac{\psi^2}{m(2m-1)}.
\end{equation}
In this region, inflationary expansion can be realized due to the false vacuum energy $V \simeq \mu^4$.
For this minimum for $\psi$, we can see
\begin{equation}
V_{,\psi\psi} \simeq \frac{8\mu^2}{(2m-1)|S|^2}
 \left[(m-1)\mu^2+
 m M^2\left(\frac{\mu^2}{m(2m-1)|S|^2}\right)^{\frac{m}{m-1}}\right] \gg \mu^4.
\end{equation}
 for $|S| \ll 1$.
Hence, we consider that the $\psi$ field can trace the instantaneous minimum 
 Eq.~(\ref{S-psi:relation}).
By substituting Eq.~(\ref{S-psi:relation}) into Eq.~(\ref{smooth:potential}),
 we obtain the potential for only 
 the canonical inflaton $\sigma = |S|/\sqrt{2}$ as
\begin{equation}
V \simeq \mu^4\left(1-\frac{2M^2}{\mu^2}\frac{m-1}{2m-1}
 \left(\frac{2\mu^2}{m(2m-1)}\right)^{\frac{m}{m-1}}\sigma^{-\frac{2m}{m-1}}
 +\frac{\sigma^4}{8}\right) .
\end{equation}
Yamaguchi and Yokoyama verified this procedure~\cite{Yamaguchi:2005qm}.
The final term in the potential comes from a supergravity effect.
The slow roll parameters are defined by
\begin{align}
\epsilon := & \frac{1}{2}\left(\frac{V_{,\sigma}}{V}\right)^2 , \\
\eta := & \frac{V_{,\sigma\sigma}}{V} ,
\end{align}
where the derivatives of the potential are given as
\begin{align}
& \frac{V_{,\sigma}}{V} \simeq \frac{2M^2}{\mu^2}\frac{2m}{2m-1}
 \left(\frac{2\mu^2}{m(2m-1)}\right)^{\frac{m}{m-1}}
 \sigma^{-\frac{2m}{m-1}-1}
 +\frac{\sigma^3}{2} , \label{smooth:V'}\\
& \frac{V_{,\sigma\sigma}}{V} \simeq -\frac{2M^2}{\mu^2}\frac{2m}{2m-1}
 \frac{3m-1}{m-1}\left(\frac{2\mu^2}{m(2m-1)}\right)^{\frac{m}{m-1}}
 \sigma^{-\frac{2m}{m-1}-2}
 +\frac{3}{2}\sigma^2 . \label{smooth:V''}
\end{align}
We define the transition point $\sigma_t$ in the potential 
 from Eq.~(\ref{smooth:V'}) as
\begin{equation}
\sigma_t^{\frac{2m}{m-1}+4} =  \frac{4M^2}{\mu^2}\frac{2m}{2m-1}
 \left(\frac{2\mu^2}{m(2m-1)}\right)^{\frac{m}{m-1}},
\end{equation} 
then Eq.~(\ref{smooth:V'}) can be recast as
\begin{equation} 
\frac{V_{,\sigma}}{V} \simeq \frac{\sigma^3}{2}\left(\frac{\sigma_t}{\sigma}\right)^{\frac{2m}{m-1}+4} +\frac{\sigma^3}{2}.
\end{equation} 
While the potential slope is dominantly given 
 by the $\sigma^{-\frac{2m}{m-1}}$ term for $\sigma < \sigma_t$, 
 the supergravity effect is significant for $\sigma > \sigma_t$.
The critical point where the slow roll condition breaks down 
 as $\eta(\sigma_c) = -1$ is given by
\begin{equation}
 \sigma_c^{\frac{2m}{m-1}+2} =
 \frac{2M^2}{\mu^2}\frac{2m}{2m-1}
 \frac{3m-1}{m-1}\left(\frac{2\mu^2}{m(2m-1)}\right)^{\frac{m}{m-1}} .
\end{equation} 
The number of e-folds during inflation is expressed as
\begin{equation}
 N = \frac{3m-1}{4m-2}\left(\frac{\sigma}{\sigma_c}\right)^{2\frac{2m-1}{m-1}},
\end{equation}
 as long as 
\begin{equation}
 \sigma(N) < \sigma_t 
\label{enq:sigma}
\end{equation}
is satisfied. 
This constrains the parameters of the model
 as shown in Fig.~\ref{Fig:m-M} where each of the contours are of $\sigma(N)/\sigma_t$
 for $N=60$ and the gray shaded region is not compatible with the condition (\ref{enq:sigma}).
If the condition (\ref{enq:sigma}) is violated, supergravity effects are not negligible
 and the resultant $n_s$ exceeds $1$~\cite{Panagiotakopoulos:1997qd,Linde:1997sj} and
 hence is not compatible with observation.

The amplitude of the density perturbation is estimated as
\begin{equation}
{\cal P}_{\zeta}
 = \frac{\mu^4}{12\pi^2}\left(\frac{m-1}{3m-1}\sigma_c\right)^{-2}
 \left(\frac{4m-2}{3m-1}N\right)^{\frac{3m-1}{2m-1}} ,
\label{eq:P}
\end{equation}
and the spectral index is expressed as
\begin{equation}
n_s = 1-\frac{3m-1}{2m-1}\frac{1}{N} .
\end{equation}
One can immediately realize that, for $m \geq 2$, the spectral index is more red tilted than that 
 of the standard supersymmetric hybrid inflation with a logarithmic correction
 in the potential.
The tensor to scalar ratio is given by
\begin{equation}
r = 16 \epsilon ,
\end{equation}
with
\begin{align}
\epsilon = & \frac{1}{2}\left(\frac{m-1}{3m-1}\left(\frac{\sigma_c}{\sigma}\right)^{\frac{3m-1}{m-1}}
 \sigma_c\right)^2 .
\end{align}
In addition, the mass of the inflaton in the vacuum is given by
\begin{equation}
m_{\sigma}^2 = 2 M^2 \left(\frac{\mu}{M}\right)^{2/m}
\left[-\mu^2+m\left(\frac{\mu}{M}\right)^{2-2/m}\right]^2 .
\label{Smooth:InflatonMass}
\end{equation}

\subsection{$m=2$ case}

\begin{figure}[htbp]
\centering
\includegraphics[clip,width=11.0cm]{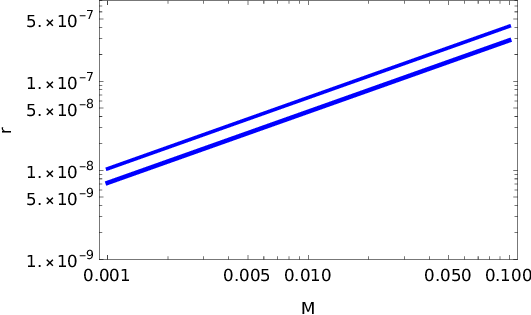}
\caption{
Cutoff scale $M$ dependence on the tensor to scalar ratio for $m=2$. 
The thicker line is for $N=60$, while the thinner line is for $N=50$.}
\label{Fig:M-r}
\end{figure}

We note some analytic formulas for $m=2$.
The condition (\ref{enq:sigma}) for a supergravity effect to be negligible
 is rewritten as
\begin{align}
1 \gg (4\sqrt{3} N^2 \mu M)^{2/3} 
 = \left(\frac{N}{55}\right)^{4/3}
 \left(\frac{\mu M}{4.8 \times 10^{-5}} \right)^{2/3} .
\label{eq:m=2:sigt}
\end{align}
The amplitude of the density perturbation reads
\begin{equation}
{\cal P}_{\zeta}
 = 20 \times 10^{-10} \left(\frac{\mu^5/M}{10^{-33/2}}\right)^{2/3}
 \left(\frac{N}{55}\right)^{5/3} ,
\label{eq:m=2:P}
\end{equation}
and the spectral index is expressed as
\begin{equation}
n_s = 1-\frac{5}{3}\frac{1}{N} \simeq 0.97 \quad {\rm for} \quad N \simeq 55.
\end{equation}
Then, from Eqs.~(\ref{eq:m=2:sigt}) and (\ref{eq:m=2:P}),
 we find that the scale of $\mu$ must be somewhat smaller than 
 a typical Grand Unified Theory (GUT) scale as $\mu < 3.4 \times 10^{-4} \simeq 8.2 \times 10^{14}$ GeV.
The tensor to scalar ratio is smaller than $\mathcal{O}(10^{-7})$ as shown in Fig.~\ref{Fig:M-r}.

\subsection{General $m$ cases}

\begin{figure}[htbp]
\centering
\includegraphics[clip,width=8.0cm]{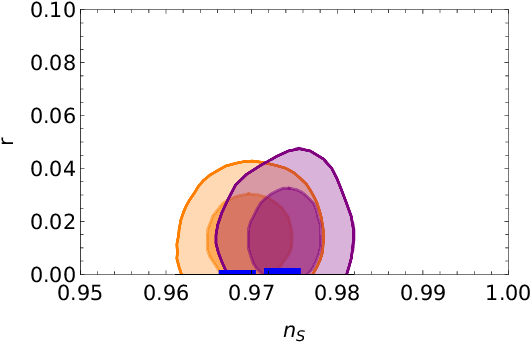}
\includegraphics[clip,width=8.0cm]{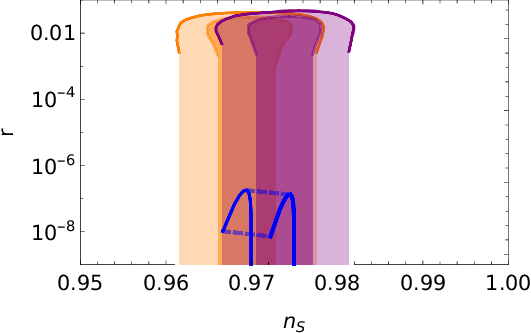}
\caption{Model prediction for $M=10^{-3}$.
The thicker blue curve is for $N=60$, while the thinner blue curve is for $N=50$.
}
\label{Fig:nsr}
\end{figure}

We show how results change if we vary $m$ in the range of $2 \leq m \leq \infty$.
The results for fixed $M=10^{-3}$ are shown in Fig.~\ref{Fig:nsr} of the $n_s-r$ plane.
The thicker (thinner) blue curve is the prediction for $N=60$ $(50)$ as $m$ varies from $2$ to $\infty$,
 where $\mu$ is adjusted to normalize the amplitude of density perturbation given in Eq.~(\ref{eq:P}).
The orangeish and purplish shadings stand for the likelihood contours
 for Planck-LB-BK18 and P-ACT-LB-BK18, respectively, reported in Ref.~\cite{ACT:2025tim}.

\subsection{Postinflationary evolution}

After inflation, the inflaton fields oscillate around the minimum (\ref{eq:minimum}).
The inflaton decay may be caused through the superpotential $W = y S H_u H_d$ 
 where $y$ is a coupling constant,
 and $H_u$ and $H_d$ are superfields of the up-type and down-type Higgs
 doublet fields, respectively.
The reheating temperature after inflation $T_R$ is controlled by the magnitude of $y$.
If $\Psi$ and $\bar{\Psi}$ are charged under $U(1)_{B-L}$,
 $\Psi$ or $\bar{\Psi}$ can couple with right-handed neutrino superfield $N$, for instance,
 as 
\begin{equation}
W = y_N \Psi NN.
\label{eq:WN}
\end{equation}
If $T_R$ is large enough, the baryon asymmetry could be generated by thermal leptogenesis~\cite{Fukugita:1986hr,Davidson:2008bu}.
On the other hand, if $T_R$ is not so high,
 the baryon asymmetry can be generated by the Affleck-Dine mechanism~\cite{Affleck:1984fy}. 
If the superpotential (\ref{eq:WN}) exists,
 nonthermal leptogenesis by the inflaton decay is also possible~\cite{Asaka:1999yd}. 
In fact, Jeannerot et al. studied nonthermal leptogenesis in
 the original smooth hybrid  inflation~\cite{Jeannerot:2001qu}.
As is suggested by the Higgs boson mass of $125$ GeV,
 soft supersymmetry (SUSY) breaking mass parameters seem to be
 large~\cite{Okada:1990vk,Okada:1990gg,Ellis:1990nz,Haber:1990aw,Ellis:1991zd}.
Thus, the promising candidate of dark matter in this scenario would be
 Higgsino-like neutralino with the mass of about $1$ TeV left due to thermal freeze-out~\cite{Cirelli:2007xd}.

\section{Conclusion}  

We revisited smooth hybrid inflation model whose
 spectral index of the density perturbation is about $0.97$,
 which well agrees with the latest results reported by ACT.
By treating the power of nonrenormalizable terms as a free parameter, 
we have found that the predicted $n_s$ lies
 within the uncertainty of the latest ACT results for any value of its power $m$. 
The scenario seems to be compatible with various baryogenesis mechanisms
 and Higgsino-like neutralino dark matter.

%
\section*{Acknowledgments}

This work is supported in part by the U.S. DOE Grants No. DE-SC0012447 and DE-
SC0023713 (N.O.) and KAKENHI Grants No. JP23K03402 (O.S.).



\end{document}